# Raman fingerprint for semi-metal WTe$_2$ evolving from bulk to monolayer

Y. C. Jiang[1], J. Gao[1, 2] & L. Wang[2, 3]

[1]*Center for Solid State Physics and Materials, School of Mathematics and Physics, Suzhou University of Science and Technology, Suzhou 215011, Jiangsu, China.*
[2]*Department of Physics, The University of Hong Kong, Pokfulam Road, Hong Kong.*
[3]*School of Materials Science and Engineering, Shanghai University, 333 Nanchen Road, Baoshan District, Shanghai 200444, China.*

**Abstract**

Tungsten ditelluride (WTe$_2$), a layered transition-metal dichalcogenide (TMD), has recently demonstrated an extremely large magnetoresistance effect, which is unique among TMDs. This fascinating feature seems to be correlated with its special electronic structure. Here, we report the observation of 6 Raman peaks corresponding to the $A_2^4$, $A_1^9$, $A_1^8$, $A_1^6$, $A_1^5$ and $A_1^2$ phonons, from the 33 Raman-active modes predicted for WTe$_2$. This provides direct evidence to distinguish the space group of WTe$_2$ from that of other TMDs. Moreover, the Raman evolution of WTe$_2$ from bulk to monolayer is clearly revealed. It is interesting to find that the $A_2^4$ mode, centered at ~109.8 cm$^{-1}$, is forbidden in a monolayer, which may be attributable to the transition of the point group from $C_{2v}$ (bulk) to $C_{2h}$ (monolayer). Our work characterizes all observed Raman peaks in the bulk and few-layer samples and provides a route to study the physical properties of two-dimensional WTe$_2$

Transition-metal dichalcogenides (TMDs) of the $MX_2$ type, where M refers to Mo and W, and X refers to S, Se and Te, have attracted much attention as graphene-like semiconductors with remarkable electric, spintronic and optical properties.[1-7] The basic structure of bulk $MX_2$ comprises two-dimensional (2D) X-M-X layers stacked through Van der Waals forces. Mono- and few-layer $MX_2$ can be fabricated by the mechanical exfoliation method.[8] The physical properties of $MX_2$ are greatly dependent on the number of layers.[9-11] Weak Van der Waals forces play a very important role in the interlayer bands and atomic vibration. As the thickness decreases, $MX_2$ exhibits a transition from indirect to direct gaps, while the frequency of the Raman peaks shift away from those of the bulk sample.[12, 13]

Tungsten ditelluride ($WTe_2$) is a unique TMD as it has the largest interlayer spacing, an orthorhombic lattice structure and a semi-metal electronic structure.[1, 14-16] This material has drawn extensive interest since the recent discovery of an extremely large magnetoresistance (XMR) effect in diamagnetic $WTe_2$ single crystal, which has not been observed in other TMDs.[1] Such a fascinating feature implies potential spintronic applications for mono- and few-layer $WTe_2$. However, to date, no study has examined the physical properties of the 2D $WTe_2$ system. Raman spectroscopy is a powerful tool for investigating the symmetries of 2D semiconductors; it is necessary, therefore, to characterize the Raman fingerprint of $WTe_2$ as it evolves from a 3D to a 2D system.

This article reports the observation of Raman scattering in bulk and few-layer $WTe_2$ samples. Six optical vibrational modes, $A_2^4$, $A_1^9$, $A_1^8$, $A_1^6$, $A_1^5$ and $A_1^2$, were observed at room temperature. Density functional perturbation theory (DFPT) was used to calculate phonon branches, simulate Raman spectra and verify their corresponding optical phonons. Group analysis demonstrated that the lattice symmetry differentiated WTe$_2$ from other TMDs.[17, 18] In our experiments, WTe$_2$ flakes, exfoliated from a piece of bulk, showed strong thickness-dependent Raman spectra. As the number of layers increased, the $A_1^8$, and $A_1^2$ modes softened (red-shift) and the $A_2^4$ and $A_1^5$ mode stiffened (blue-shift), suggesting that not only the interlayer Van der Waals coupling, but also stacking-induced structural changes or Coulombic interlayer interaction, affected the atomic vibration. It was also interesting to find that the $A_2^4$ mode was absent in the monolayer. This phenomenon may be attributable to the transition of the point group from $C_{2v}$ to $C_{2h}$. Our results provide a well-defined and reliable method for determining the number of layers using Raman spectroscopy.

**Results and discussions**

Fig. 1a shows the layered crystal structure of WTe$_2$, in which tungsten layers are sandwiched between two neighboring tellurium sheets. The distance (0.703 nm) between two adjacent sandwich layers is the largest of all known TMDs. Most TMDs exhibit a hexagonal structure with the space group $D^4_{6h}$. However, naturally formed WTe$_2$ reveals an orthorhombic unit cell with the space group $C^7_{2v}$ (*Pmn*2$_1$), which may be derived from a distorted hexagonal net.[1, 15] The unit cell contains two tungsten atoms and four tellurium atoms. The two tungsten atoms (distinguished by W1 and W2) are not equivalent to each other and the W-Te bond length ranges from 2.7 to 2.8 Å. WTe$_2$ bulk exhibits the T$_d$ stacking order, in which the atoms in the upper layer are rotated by 180 degrees with respect to those in the lower layer.[17] The tungsten atoms also shift away from the center of the octahedron formed by the tellurium atoms. Therefore, the vibrational modes of Td-WTe$_2$ are very different from those of 2H-MX$_2$, and the previous calculation of vibrational modes, based on 2H-WTe$_2$, is inapplicable to our experimental Raman results.[19]

Due to the $C^7_{2v}$ symmetry group, the irreducible representations of the optical phonons in the bulk WTe$_2$ at the center of the Brillouin zone ($\Gamma$ point) are

$$\Gamma_{\text{bulk}} = 11A_1 + 6A_2 + 11B_1 + 5B_2,$$

where all of the vibrational modes are Raman active, and the 11A$_1$, 11B$_1$ and 5B$_2$ modes are infrared active. Here, the Raman tensors corresponding to different symmetries can be written as

$$\alpha(A_1) = \begin{pmatrix} a & 0 & 0 \\ 0 & b & 0 \\ 0 & 0 & c \end{pmatrix}, \quad \alpha(A_2) = \begin{pmatrix} 0 & d & 0 \\ d & 0 & 0 \\ 0 & 0 & 0 \end{pmatrix},$$

$$\alpha(B_1) = \begin{pmatrix} 0 & 0 & 0 \\ 0 & 0 & e \\ 0 & e & 0 \end{pmatrix}, \quad \alpha(B_2) = \begin{pmatrix} 0 & 0 & f \\ 0 & 0 & 0 \\ f & 0 & 0 \end{pmatrix}.$$

Although 33 Raman vibrations are predicted by group theory, only 6 Raman peaks were observed in our experiments. To distinguish these vibrational modes, we use $Z^n_m$ (Z = A or B; m = 1 or 2; n = integer) to represent them. The frequency of $Z^n_m$ mode is smaller than that of $Z^l_m$ mode if $n$ is larger than $l$. This method allows us to name all appropriate vibrational modes.

DFPT has been put into use to calculate all 33 Raman vibrational modes. Among them, the atomic displacements of the six Raman active modes that could potentially explain our Raman results are shown in Fig. 1b. Compared with the Raman active modes of 2H-MX$_2$, two non-equivalent W atoms (W1 and W2) are key factors to determine the anomalous Raman vibrations of WTe$_2$.

The Raman spectra of WTe$_2$ bulk at room temperature are presented in Fig. 2a. Raman peaks centered at ~109.8 cm$^{-1}$, ~130.4 cm$^{-1}$ ~162.5 cm$^{-1}$ and 208.4 cm$^{-1}$ were observed with Laser I applied in the $z(xx)\bar{z}$ geometry. Another two Raman peaks, centered at ~119.6 cm$^{-1}$ and ~139.1 cm$^{-1}$, were observed with Laser II applied in the $y(zz)\bar{y}$ geometry. . Due to the non-polarized Raman measurement, the $\vec{e}_i$ and $\vec{e}_s$ vectors are not along a- or c- axis in the either geometry. Thus, a total of six Raman peaks were identified in our experiments. Due to the non-polarized Raman measurements, the assignment of these Raman peaks could not be performed through symmetry analysis. Instead, we calculated the Raman intensities and spectrum (green curve in Fig. 2a) to verify the phonon modes corresponding to the observed Raman peaks. Only 6 Raman peaks were observed in comparison with 33 Raman modes in theory, which suggests that only those modes with high Raman intensities might appear experimentally. For example, Table 1 shows that for the first Raman peak (centered at ~109.8 cm$^{-1}$), the $A_2^4$, $A_2^5$ and $B_2^4$ modes were all candidates, with frequencies close to ~110 cm$^{-1}$. Only the $A_2^4$ mode exhibited a sufficiently strong Raman intensity, whereas the intensities of the others were very weak. Therefore, the $A_2^4$ mode should be assigned to the first peak. Similarly, the $A_1^9$ (~119.6 cm$^{-1}$), $A_1^8$ (~130.4 cm$^{-1}$), $A_1^6$ (~139.1 cm$^{-1}$), $A_1^5$ (~162.5 cm$^{-1}$) and $A_1^2$ (~208.4 cm$^{-1}$) modes were assigned frequencies from low to high, which agrees with the results in Ref. 21. It should be noted that although the $A_1^2$ mode was associated with the peak centered at ~208.4 cm$^{-1}$, the $A_1^3$ mode still exhibited a strong Raman intensity and could not be neglected. It is very possible that the $A_1^2$ and $A_1^3$ modes both exist, but the difference between their frequencies is so small that they cannot be distinguished in the Raman spectra.

As shown in Fig. 1b, the first five modes originate from the relative movements of Te atoms, while the $A_1^2$ mode is due to the displacement between adjacent W1 and W2 atoms. $A_2^4$ and $A_1^6$ are pure longitudinal and transverse optical modes, respectively. For the in-plane configuration [$z(xx)\bar{z}$], the B$_1$ and B$_2$ modes are forbidden due to the forms of their Raman tensors. In the case of the out-of-plane configuration [$y(zz)\bar{y}$], the $A_2^4$ mode may be forbidden and the $A_1^8$ Raman intensity vanishes. Instead, the $A_1^6$ and $A_1^9$ modes reveal significant Raman intensities. Here when we state that one mode is "forbidden", it means that a Raman mode is prevented from observing by group theory or unable to exist on account of the DFPT calculation, while "vanish" means a Raman mode is active in theory, but in a certain configuration its intensity is too weak to observe due to the specific parameters in its Raman tensor. To clarify this phenomenon, the relationship between Raman intensity and Raman tensor can be written as[20]

$$I = [e_s \alpha e_i]^2 |E_0|^2,$$

where $e_s$ and $e_i$ are the scattered light and incident light vectors, respectively. This equation makes it easy to understand the anisotropy of the $A_2^4$ mode. The specific values of a, b and c in the Raman tensors of the $A_1^8, A_1^6$ and $A_1^9$ modes determine the dependence on the laser configuration. In the case of Laser I, the $A_1^6$ and $A_1^9$ modes with very small values of a and b vanish, while the $A_1^8$ mode appears due to its large value of b. There is a similar effect for Laser II: the $A_1^6$ and $A_1^9$ modes appear due to their large c values and the $A_1^8$ mode vanishes due to its very small a and c values. The a, b and c values of the $A_1^2$ and $A_1^5$ modes are large enough for them to be observed in both configurations.

Fig. 2b shows the calculated phonon dispersion curve in the Brillouin zone (BZ), and DOS for WTe$_2$ bulk. There are three acoustic (green curves) and thirty-three optical (blue curves) phonon branches. Table 1 shows all of the calculated and experimental vibrational modes. The $A_2^4$, $A_1^9$, $A_1^8$, $A_1^6$, $A_1^5$ and $A_1^2$ modes are calculated to have strong Raman intensities at frequencies of ~110.0 cm$^{-1}$, ~119.4 cm$^{-1}$, ~130.0 cm$^{-1}$, ~142.3 cm$^{-1}$, ~168.5 cm$^{-1}$, and 222.2 cm$^{-1}$, respectively. The good agreement between experimental and theoretical results supports our assignments of the Raman vibrational modes. Moreover, it is necessary to note that all calculations were done at zero temperature, while the Raman spectra were investigated at room temperature. However, the effect of temperature cannot influence the assignment of Raman modes. The frequency displacements between 5 K and 294 K are less than 4 cm$^{-1}$ for all the observed Raman peaks.[21] Such small displacements would not misguide the association with specific phonon modes. In addition, all Raman peaks shift up in frequency with the temperature decreasing, which may explain why most of the calculated frequencies are higher than those measured in experiments.

WTe$_2$ flakes were fabricated on Si substrate with 300 nm SiO$_2$ layer using the mechanical exfoliation method. As shown in Fig. 3a and b, an ultrathin WTe$_2$ flake were first identified by an optical microscopy, and then imaged by AFM to determine their thickness. Fig. 3c shows that a monolayer WTe$_2$ on SiO$_2$ layer has the height of about 1.1 nm, while the height of monolayer on the WTe$_2$ flake is 0.7~0.9 nm and close to the interlayer spacing (about 0.703 nm) of bulk.

Fig. 4a illustrates Raman spectra of WTe$_2$ evolving from bulk to monolayer. The four Raman active modes can be observed in 2L, 3L, 4L and 5L samples. It is interesting to note that the $A_2^4$ mode absent in the 1L sample, which may make it convenient to identify the WTe$_2$ monolayer through Raman spectra. The frequency difference between the $A_2^4$ modes of the 2L sample and bulk is about 2.7 cm$^{-1}$. The $A_2^4$ and $A_1^5$ modes stiffen (blue-shifts) in frequency as the layer number increases, whereas the $A_1^8$ and $A_1^2$ modes soften (red-shift). The $A_1^5$ mode is the most stable and the displacement of its Raman shift, caused by the changing of thickness, is less than 1 cm$^{-1}$. In contrast, the $A_1^2$ mode is found to the most sensitive to the layer number, and shifts up by about 4.8 cm$^{-1}$ with the layer number decreasing to one.

With WTe$_2$ evolving from bulk to monolayer, its point group changes from $C_{2v}$ to $C_{2h}$. The irreducible representation of the optical phonons in a WTe$_2$ monolayer at the Brillion zone center ($\Gamma$ point) is expressed as:

$$\Gamma_{monolayer} = 6A_g + 2A_u + 3B_g + 4B_u,$$

where $A_g$ and $B_g$ are Raman active; $A_u$ and $B_u$ are infrared active. Only nine Raman-active modes are allowed in the WTe$_2$ monolayer. This implies that some Raman peaks, observed in bulk, may be forbidden in the monolayer. Based on the group analysis, we calculated phonon dispersion curve in BZ, and DOS for WTe$_2$ monolayer as shown in Fig. 4b. There are three acoustic (green curves) and fifteen optical (blue curves) phonon branches.

Vibrational modes of the WTe$_2$ monolayer are exhibited in Table 2. To compare the vibrational modes of bulk with those of monolayer, we use the symmetry symbols of both $C_{2v}$ and $C_{2h}$ groups to name the corresponding modes. The atomic displacements of the Raman active modes were used to build a bridge between Raman modes of monolayer and bulk WTe$_2$ as shown in Fig. 1b. For example, the $A_1^5$ and $B_1^5$ modes correspond to the $A_g^3$ mode in Table 2, which indicates that the $A_1^5$ and $B_1^5$ modes have the same atomic displacement as that of the $A_g^3$ mode. In the $A_1^5$ mode, the upper and lower atom layers exhibit the same phase of the atomic oscillations. In the $B_1^5$ mode, the upper atom layer has an anti-phase with respect to the lower atom layer. Both of the modes can be distinguished in the multilayer samples (layer number⩾2), but turn into the $A_g^3$ mode in the monolayer.

It is found that the $A_2^4$ mode is forbidden in the monolayer, which explains the absence of $A_2^4$ Raman peak. Fig. 4c shows the calculated Raman spectrum of the WTe$_2$ monolayer. Only the $B_g^2$ mode is a candidate to correspond to the $A_2^4$ mode. However, Fig. 1b demonstrates that the atomic displacement of the $B_g^2$ mode is different from that of the $A_2^4$ mode, but the same as that of the $B_2^4$ mode. Therefore, the $B_g^2$ mode corresponds to the $B_2^4$ mode instead of the $A_2^4$ mode. On the basis of the theoretical analysis, the transition of point group from $C_{2v}$ (bulk) to $C_{2h}$ (monolayer) may be responsible for the absence of the $A_2^4$ mode. In addition, the atomic displacement of the $A_2^4$ mode shows that only the half of tellurium atoms participate in oscillation, while the other half remains still. This phenomenon may be attributed to the interlayer Van der Waals force. With WTe$_2$ bulk evolving into monolayer, all of the tellurium atoms have to participate in oscillation due to lack of the interlayer Van der Waals force.

The frequencies of the four modes as functions of the layer number are shown in Fig. 4d. It is observed that the $A_1^8$, and $A_1^2$ modes soften and the $A_1^5$ and $A_2^4$ modes stiffen as WTe$_2$ monolayer evolves into bulk. This implies that not only the interlayer Van der Waals coupling, but also stacking induced structure changes or Coulombic interlayer interaction can impact on the atomic vibration.[22] Our calculated results of the WTe$_2$ monolayer confirmed shift directions of these vibrational modes.

In summary, bulk and few-layer WTe$_2$ have been studied using Raman spectroscopy. The six first-order Raman-active modes were observed at room temperature. DFPT and group theory were used to analyze frequencies of the Raman peaks and their corresponding vibration modes. It was found that the Raman peaks were correlated with the optical phonons of $A_2^4$, $A_1^9$, $A_1^8$, $A_1^6$, $A_1^5$ and $A_1^2$. The $A_2^4$ mode was found to be forbidden in the monolayer due to the transition of point group from $C_{2v}$ (bulk) to $C_{2h}$ (monolayer). With the layer number increasing, the $A_1^8$, and $A_1^2$ modes softened and the $A_2^4$ and $A_1^5$ modes stiffened. Our work characterized all Raman peaks observed in bulk and few-layer $WTe_2$ and provided a route to study the physical properties of 2D $WTe_2$.

**Methods**

The $WTe_2$ single crystals were prepared using a chemical vapor transport method described by Ref.1. $WTe_2$ flakes were mechanically exfoliated from a piece of bulk single crystal onto Si wafers covered with a 300-nm-thick $SiO_2$ layer. The few-layer samples were first identified by optical microscopy, and then measured by atomic-force microscopy (AFM) to determine the thickness. Micro-Raman spectroscopy was used to analyze the $WTe_2$ under ambient conditions. Non-polarized and off-resonance Raman measurements were performed with a 514.5-nm excitation laser. When the $WTe_2$ flakes were investigated, the power of the excitation laser was reduced to 0.5 mW to obtain reliable Raman spectra against the oxidation. Our calculation and analysis were performed on basis of DFPT and group theory, using the experimental lattice parameters a=3.496 Å, b=6.282 Å and c=14.07 Å and the same wavelength of the incident light as that in experiments.[14] The exchange correlation potential was represented by the local density approximation (LDA). The Brillouin zone integrations were performed with a $16\times8\times4$ Monkhorst-Pack k-point mesh by using a plane-wave energy cutoff of 500 eV. All DFPT calculations model was appropriate for explaining the off-resonance Raman spectrum using the CASTEP code.[23]

**Reference**

1.  Ali, M. N. *et al.* Large, non-saturating magnetoresistance in WTe$_2$. *Nature* **514**, 205-208 (2014).

2.  Radisavljevic, B., Radenovic, A., Brivio, J., Giacometti, V. & Kis, A. Single-layer MoS$_2$ transistors. *Nat. Nano.* **6**, 147-150 (2011).

3.  Coleman, J. N. *et al.* Two-dimensional nanosheets produced by liquid exfoliation of layered materials. *Science* **331**, 568-571 (2011).

4.  Geim, A. K. & Grigorieva, I. V. Van der Waals heterostructures. *Nature* **499**, 419-425 (2013).

5.  Gong, Y. *et al.* Vertical and in-plane heterostructures from WS$_2$/MoS$_2$ monolayers. *Nat. Mater.* **13**, 1135-1142 (2014).

6.  Zeng, H., Dai, J., Yao, W., Xiao, D. & Cui, X. Valley polarization in MoS$_2$ monolayers by optical pumping. *Nat. Nano.* **7**, 490-493 (2012).

7.  Zeng, H. *et al.* Optical signature of symmetry variations and spin-valley coupling in atomically thin tungsten dichalcogenides. *Sci. Rep.* **3**, 1608 (2013).

8.  Novoselov, K. S. *et al.* Two-dimensional atomic crystals. *Proc. Natl. Acad. Sci. U.S.A.* **102**, 10451-10453 (2005).

9.  Lee, C., Yan, H., Brus, L. E., Heinz, T. F., Hone, J. & Ryu, S. Anomalous Lattice Vibrations of Single- and Few-Layer MoS$_2$. *ACS Nano* **4**, 2695-2700 (2010).

10. Li, H. *et al.* From Bulk to Monolayer MoS$_2$: Evolution of Raman Scattering. *Adv. Funct. Mater.* **22**, 1385-1390 (2012).

11. Yamamoto, M. *et al.* Strong Enhancement of Raman Scattering from a Bulk-Inactive Vibrational Mode in Few-Layer MoTe$_2$. *ACS Nano* **8**, 3895-3903 (2014).

12. Mak, K. F., Lee, C., Hone, J., Shan, J. & Heinz, T. F. Atomically Thin MoS$_2$: A New Direct-Gap Semiconductor. *Phys. Rev. Lett.* **105**, 136805 (2010).

13. Splendiani, A. *et al.* Emerging Photoluminescence in Monolayer MoS$_2$. *Nano Lett.* **10**, 1271-1275 (2010).

14. Brown, B. The crystal structures of WTe$_2$ and high-temperature MoTe$_2$. *Acta*

*Crysta.* **20**, 268-274 (1966).

15. Mar, A., Jobic, S. & Ibers, J. A. Metal-metal vs tellurium-tellurium bonding in $WTe_2$ and its ternary variants $TaIrTe_4$ and $NbIrTe_4$. *J. Am. Chem. Soc.* **114**, 8963-8971 (1992).

16. Kabashima, S. Electrical Properties of Tungsten-Ditelluride $WTe_2$. *J. Phys. Soc. Jpn.* **21**, 945-948 (1966).

17. Augustin, J. *et al.* Electronic band structure of the layered compound Td-$WTe_2$. *Phys. Rev. B* **62**, 10812-10823 (2000).

18. Rousseau, D. L., Bauman, R. P. & Porto, S. P. S. Normal mode determination in crystals. *J. Raman Spectrosc.* **10**, 253-290 (1981).

19. Ataca, C., Şahin, H. & Ciraci, S. Stable, Single-Layer $MX_2$ Transition-Metal Oxides and Dichalcogenides in a Honeycomb-Like Structure. *J. Phys. Chem. C* **116**, 8983-8999 (2012).

20. Bower, D. I. Investigation of Molecular Orientation Distributions by Polarized Raman Scattering and Polarized Fluorescence. J. Polym. Sci.:Polym. Phys. Ed. 10, 2135-2153 (1972)

21. Kong, W.-D. *et al*. Raman scattering investigation of large positive magnetoresistance material $WTe_2$. *Appl. Phy. Lett.* **106**, 081906 (2015).

22. Wieting, T. J. & Verble, J. L. Interlayer Bonding and the Lattice Vibrations of β-GaSe. *Phys. Rev. B* **5**, 1473-1479 (1972).

23. Clark, S. J. *et al*. First principles methods using CASTEP. *Z. Kristallogr* **220**, 567 (2005).

**Acknowledgements**

The authors thank Dr. An Zhao, Dr. Bairen Zhu and Dr. Xiangbo Liu for their fruitful discussions, and Miss Manxiu Feng for her technique support. This work has been supported by the National Key Project for Basic Research (No. 2014CB921002), the National Natural Science Foundation of China (Grant No. 11504254, 11374225, 11574227), and the Research Grant Council of Hong Kong (Project code:701813). This work is also supported by the Priority Academic Program Development of Jiangsu Higher Education Institutions and the USTS Cooperative Innovation Center for Functional Oxide Films and Optical Information.

**Author contributions**

J.G. proposed and led the project. Y.C.J. designed and did all the experiments. Y.C.J. and L.W. performed the DFPT calculations. Y.C.J. and J.G. wrote the manuscript and prepared all figures together.

**Additional information**

Competing financial interests: The authors declare no competing financial interests.

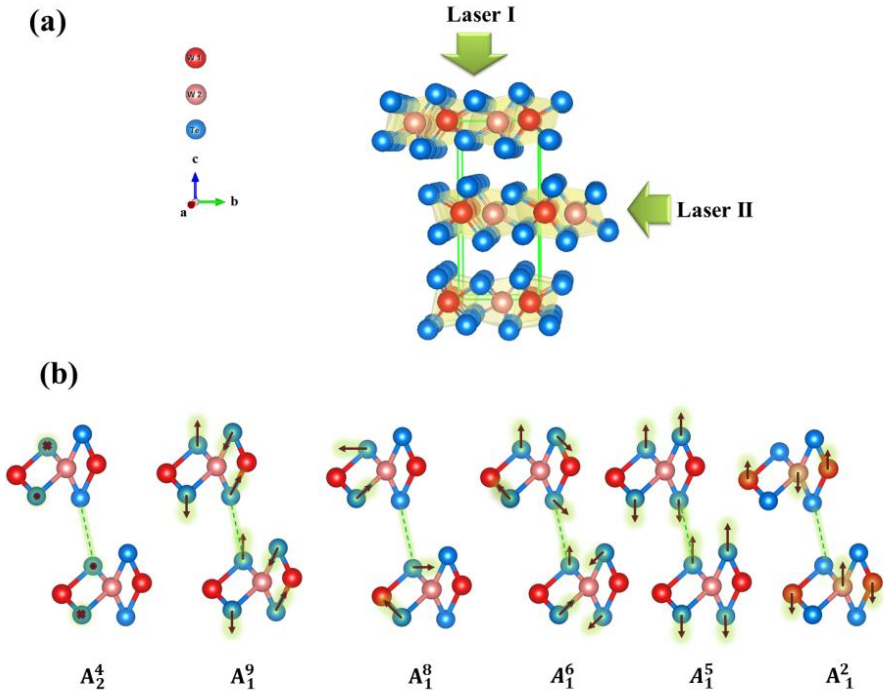

**Figure 1| Lattice structure** (a) Crystal structure of WTe$_2$ along the a-axis direction, showing the orthorhombic unit cell. (b) Atomic displacements of Raman active modes existing in WTe$_2$ bulk and monolayer. Here, "×" and "·" mean that Te atoms move into and out of the *bc* plane, respectively.

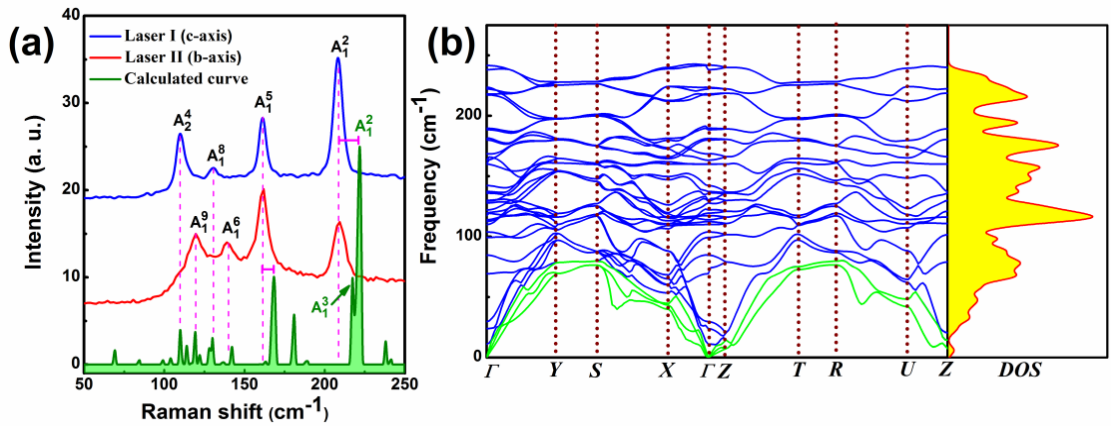

**Figure 2| Experimental and calculated phonons** (a) Raman spectra of the bulk sample with incident Laser I and Laser II along *c*- and *b*-axes at room temperature, respectively. Calculated Raman spectrum is plotted for comparison. (b) Calculated phonon dispersion curve along the *Γ-Y-X-Γ-Z* direction in the orthorhombic Brillouin

zone, and vibrational density of states (DOS) for WTe$_2$ bulk at the equilibrium volume. Blue and green curves are optical and acoustic vibrational branches, respectively.

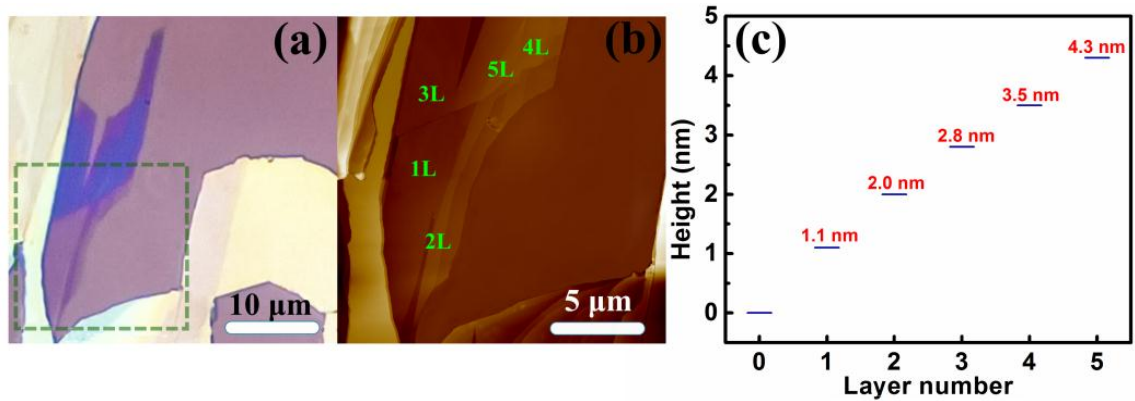

**Figure 3| Microscopic Imaging** (a) Optical microscopic images of mono- to multilayer WTe$_2$ on Si substrate with 300 nm SiO$_2$ layer. (b) An AFM image of the area (20×20 μm$^2$) surrounded by green dashed line in (a). (c) Height of the flake as a function of the layer number.

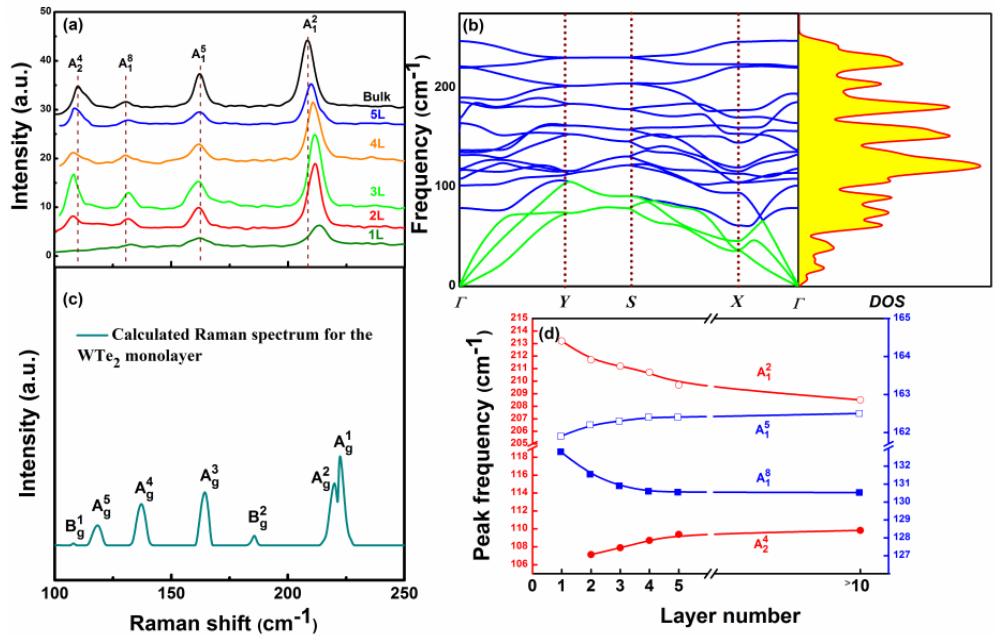

**Figure 4| Raman evolution** (a) Raman spectra of bulk and few-layer WTe$_2$. (b) Calculated phonon dispersion curve along the $\Gamma$-$Y$-$X$-$\Gamma$ direction in the orthorhombic BZ, and DOS for WTe$_2$ monolayer. Blue and green curves are optical and acoustic

vibrational branches, respectively. (c) Calculated Raman spectrum for the WTe$_2$ monolayer. (d) Frequencies of the $A_2^4$, $A_1^8$, $A_1^5$ and $A_1^2$ modes as functions of the layer number.

| Symmetry | Raman Activity | Infrared Activity | Experimental Frequency (cm$^{-1}$) | Calculated Frequency (cm$^{-1}$) |
|---|---|---|---|---|
| $A_1^{11}$ | Y | Y | | 10.9 |
| $B_1^{11}$ | Y | Y | | 23.9 |
| $A_2^6$ | Y | N | | 29.7 |
| $A_1^{10}$ | Y | Y | | 69.3 |
| $B_1^{10}$ | Y | Y | | 84.8 |
| $B_2^5$ | Y | Y | | 99.3 |
| $A_2^5$ | Y | N | | 103.9 |
| $A_2^4$ | Y | N | 109.8 | 110.0 |
| $B_2^4$ | Y | Y | | 112.5 |
| $A_2^3$ | Y | N | | 114.0 |
| $B_2^3$ | Y | Y | | 117.4 |
| $A_1^9$ | Y | Y | 119.6 | 119.4 |
| $B_1^9$ | Y | Y | | 122.1 |
| $B_1^8$ | Y | Y | | 127.9 |
| $B_1^7$ | Y | Y | | 129.5 |
| $A_1^8$ | Y | Y | 130.4 | 130.0 |
| $A_1^7$ | Y | Y | | 136.0 |
| $B_1^6$ | Y | Y | | 137.3 |
| $A_1^6$ | Y | Y | 139.1 | 142.3 |
| $A_2^2$ | Y | N | | 160.0 |
| $B_2^2$ | Y | Y | | 160.1 |
| $B_1^5$ | Y | Y | | 163.3 |
| $A_1^5$ | Y | Y | 162.5 | 168.5 |
| $A_2^1$ | Y | N | | 180.1 |
| $B_2^1$ | Y | Y | | 181.2 |
| $B_1^4$ | Y | Y | | 188.6 |
| $A_1^4$ | Y | Y | | 189.7 |
| $A_1^3$ | Y | Y | | 217.3 |
| $B_1^3$ | Y | Y | | 218.1 |
| $A_1^2$ | Y | Y | 208.4 | 222.2 |
| $B_1^2$ | Y | Y | | 223.1 |

| Symmetry | | | 238.1 |
|---|---|---|---|
| $B_1^1$ | Y | Y | 238.1 |
| $A_1^1$ | Y | Y | 241.7 |

**Table 1.** All of the possible optical phonons investigated through calculations and experiments. "Y" or "N" means that the chosen mode is allowed or forbidden.

| Symmetry | Symmetry (bulk) | Experimental Frequency (cm$^{-1}$) | Calculated Frequency (cm$^{-1}$) |
|---|---|---|---|
| $A_g^6$ | $A_1^{10}$ or $B_1^{10}$ | | 81.4 |
| $B_g^3$ | $A_2^5$ or $B_2^5$ | | 93.2 |
| $B_g^2$ | $B_2^4$ | | 108.5 |
| $A_u^2$ | $A_2^3$ or $B_2^3$ | | 114.3 |
| $A_g^5$ | $A_1^9$ | | 118.5 |
| $B_u^4$ | $B_1^8$ | | 132.2 |
| $B_u^3$ | $A_1^7$ or $B_1^7$ | | 135.1 |
| $A_g^4$ | $A_1^8$ or $B_1^6$ | 132.8 | 137.2 |
| $A_u^1$ | $A_2^2$ or $B_2^2$ | | 156.6 |
| $A_g^3$ | $A_1^5$ or $B_1^5$ | 161.9 | 164.5 |
| $B_g^1$ | $A_2^1$ or $B_2^1$ | | 185.7 |
| $B_u^2$ | $A_1^4$ or $B_1^4$ | | 190.2 |
| $A_g^2$ | $A_1^3$ | | 220.3 |
| $A_g^1$ | $A_1^2$ | 213.2 | 222.6 |
| $B_u^1$ | $A_1^1$ or $B_1^1$ | | 247.2 |

**Table 2.** The calculated vibrational modes of the WTe$_2$ monolayer.